\documentclass[fleqn,usenatbib]{mnras}

\usepackage{newtxtext,newtxmath}
\usepackage{hyperref}
\usepackage{multirow}
\usepackage{tabularx}
\usepackage{xcolor}
\usepackage{bm}

\usepackage[T1]{fontenc}

\DeclareRobustCommand{\VAN}[3]{#2}
\let\VANthebibliography\thebibliography
\def\thebibliography{\DeclareRobustCommand{\VAN}[3]{##3}\VANthebibliography}

\usepackage{graphicx}	
\usepackage{amsmath}	

\newcommand{\revone}[1]{#1} 

\title[TRON II. A radio flare from an RS CVn]{Mining the time axis with TRON. II. MeerKAT detects a stellar radio flare from a distant RS CVn candidate}

\author[Smirnov, O.M. et al.]{O.~M.~Smirnov$^{1,2,3}$\thanks{E-mail: o.smirnov@ru.ac.za},
A.~Golden$^{4}$, T.~Myburgh$^{1}$, B.~Ngcebetsha$^{2,1}$, 
C.~Tasse$^{5,1}$, I.~Heywood$^{6,7,1,2}$, \newauthor
A.~J.~T.~Ramaila$^{2,1}$, M.~A.~Thompson$^{8}$, 
J.~S.~Kenyon$^{1}$, S.~J.~Perkins$^{2}$, 
J.~Dawson$^{1,2}$, H.~L.~Bester$^{2,1}$, \newauthor
J.~S.~Bright$^{6,7}$, N.~Oozeer$^{2,1}$,
V. G. G.~Samboco$^{1}$, I.~Sihlangu$^{2,1}$, C.~Choza$^{6,7}$
\\
$^{1}$Centre for Radio Astronomy Techniques and Technologies (RATT), Department of Physics and Electronics, Rhodes University, Makhanda, 6140,\\South Africa\\
$^{2}$South African Radio Astronomy Observatory, Cape Town, 7925, South Africa\\
$^{3}$Institute for Radioastronomy, National Institute of Astrophysics (INAF IRA), Via Gobetti 101, 40129 Bologna, Italy \\
$^{4}$Physics, School of Natural Sciences, Ollscoil na Gaillimhe --- University of Galway, University Road, Galway, H91 TK33, Ireland \\
$^{5}$GEPI \& ORN, Observatoire de Paris, Université PSL, CNRS, 5 Place Jules Janssen, 92190 Meudon, France\\
$^{6}$Astrophysics, Department of Physics, University of Oxford, Keble Road, Oxford, OX1 3RH, UK\\
$^{7}$Breakthrough Listen, Astrophysics, Department of Physics, The University of Oxford, Keble Road, Oxford, OX1 3RH, UK\\
$^{8}$School of Physics and Astronomy, University of Leeds, Woodhouse Lane, Leeds, LS2 9JT, UK\\
}

\date{Accepted X. Received Y; in original form Z}

\pubyear{2024}

\begin{document}

\label{firstpage}
\pagerange{\pageref{firstpage}--\pageref{lastpage}}
\maketitle

\begin{abstract}

Medium-timescale (minutes to hours) radio transients are a relatively unexplored population. The wide field-of-view and high instantaneous sensitivity of instruments such as MeerKAT provides an opportunity to probe this class of sources, using image-plane detection techniques. The previous letter in this series describes our project and associated TRON pipeline designed to mine archival MeerKAT data for transient and variable sources. In this letter, we report on a new transient, a radio flare, associated with Gaia DR3 6865945581361480448, a G type star, whose parallax places it at a distance of 1334~pc. Its duration and high degree of circular polarization suggests electron cyclotron maser instability as the mechanism, consistent with an RS CVn variable.

\end{abstract}

\begin{keywords}
   radio continuum: transients -- stars -- 
   methods: data analysis -- techniques: interferometric
\end{keywords}

\section{Introduction} 

MeerKAT's high instantaneous sensitivity and large field of view make it an excellent instrument for detecting radio transients. The \revone{first letter of this series\footnote{\url{https://arxiv.org/abs/2501.09488}} \citep{mining1}} introduced our project of mining archival MeerKAT data for new transients and variables using high time cadence interferometric imaging, briefly discussed the TRON pipeline developed for this project, and reported initial detections of millisecond pulsars in globular clusters.

In this letter, we report on a new transient source detected by TRON. The source has a clear stellar origin. The detection was made in a single pointing that was part of the PARROT transient follow-up \citep{parrot}, thus making for two completely unrelated (and astrophysically very different) transient detections in a single MeerKAT observation, which underscores the opening point of our introduction.

\section{Observations and data reduction}

The observation reported on here were conducted by MeerKAT in conventional synthesis imaging mode, using the L-band (856--1712 MHz) and UHF (544--1088 MHz) systems in 4096 channel correlator mode, with 8~s integrations.

The detection observation was a 10-hour L-band track taken on 20 June 2021, consisting of conventional tracking scans on target, interspersed with shorter scans on a bandpass and gain calibrator. Details of all observations are given in \citet{parrot}; this particular epoch is referred to as ``L2'' therein. Other observations reported on here were structured similarly, using a mix of L-band and UHF systems.

\paragraph*{Data reduction.} 

As highlighted in the first letter of this series, the inputs to TRON are, generally, yielded by any conventional calibration and imaging workflow. All that is required is a  set of calibrated visibilities, and a deep sky model. For the PARROT observations, we used the calibration pipeline described by \citet{parrot}. The corrected data and model image from the pipeline was then passed through TRON, as described in Letter I \citep{mining1}.  

\section{Results}

\begin{figure}
   \includegraphics[width=\columnwidth]{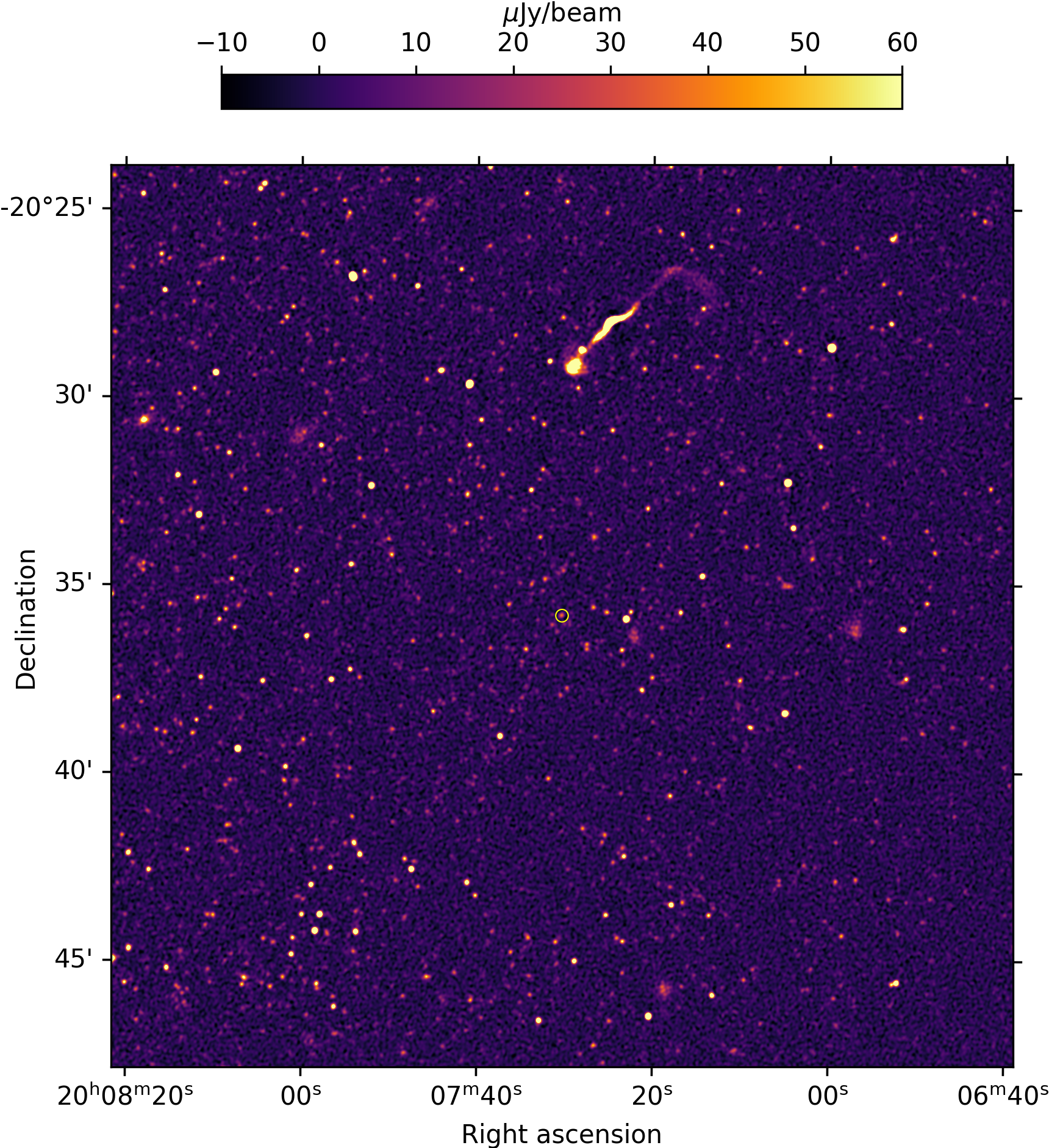}
   \caption{\label{fig:l2}MeerKAT image of the field surrounding Gaia DR3 6865945581361480448 at 1.28 GHz. The position of the star is marked by a yellow circle.}
\end{figure}
  
\begin{figure}
   \includegraphics[width=\columnwidth]{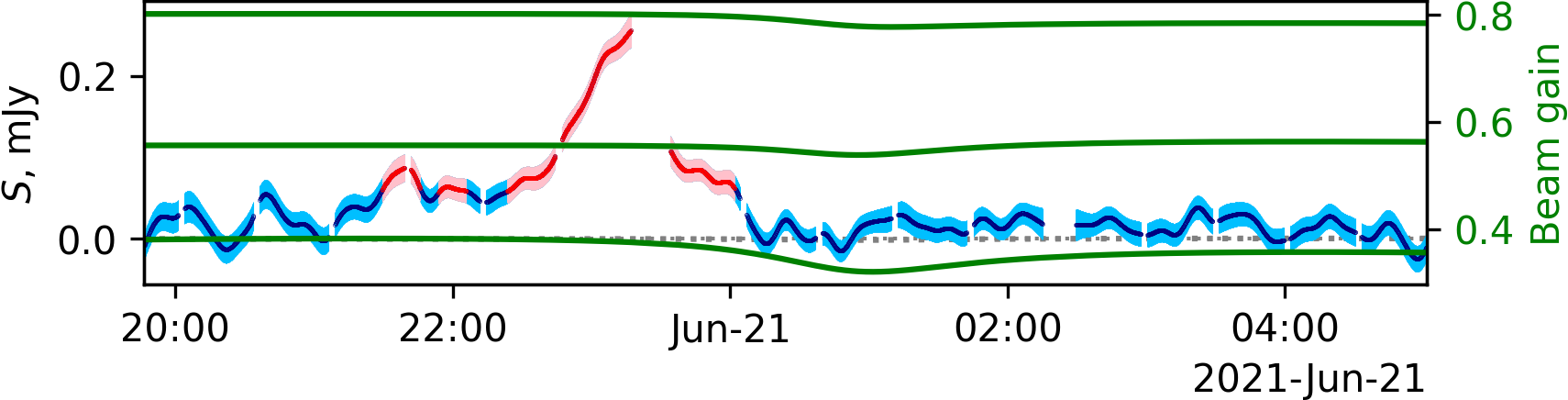}
   \caption{\label{fig:l2-lc}Full L-band, 240s-smoothed lightcurve of Gaia DR3 6865945581361480448. Units are apparent flux. \revone{Time is UTC.} Error bars (computed as the local image rms) are plotted in light blue. Four-sigma deviations are indicated in red, with light red error bars.
   The green curves (refer to the right $y$ axis for scale), show the power beam gain in the direction of the source as a function of time for the bottom (highest gain), mid and top (lowest gain) end of the band.}
\end{figure}
 

\begin{figure*}
   \includegraphics[width=\textwidth]{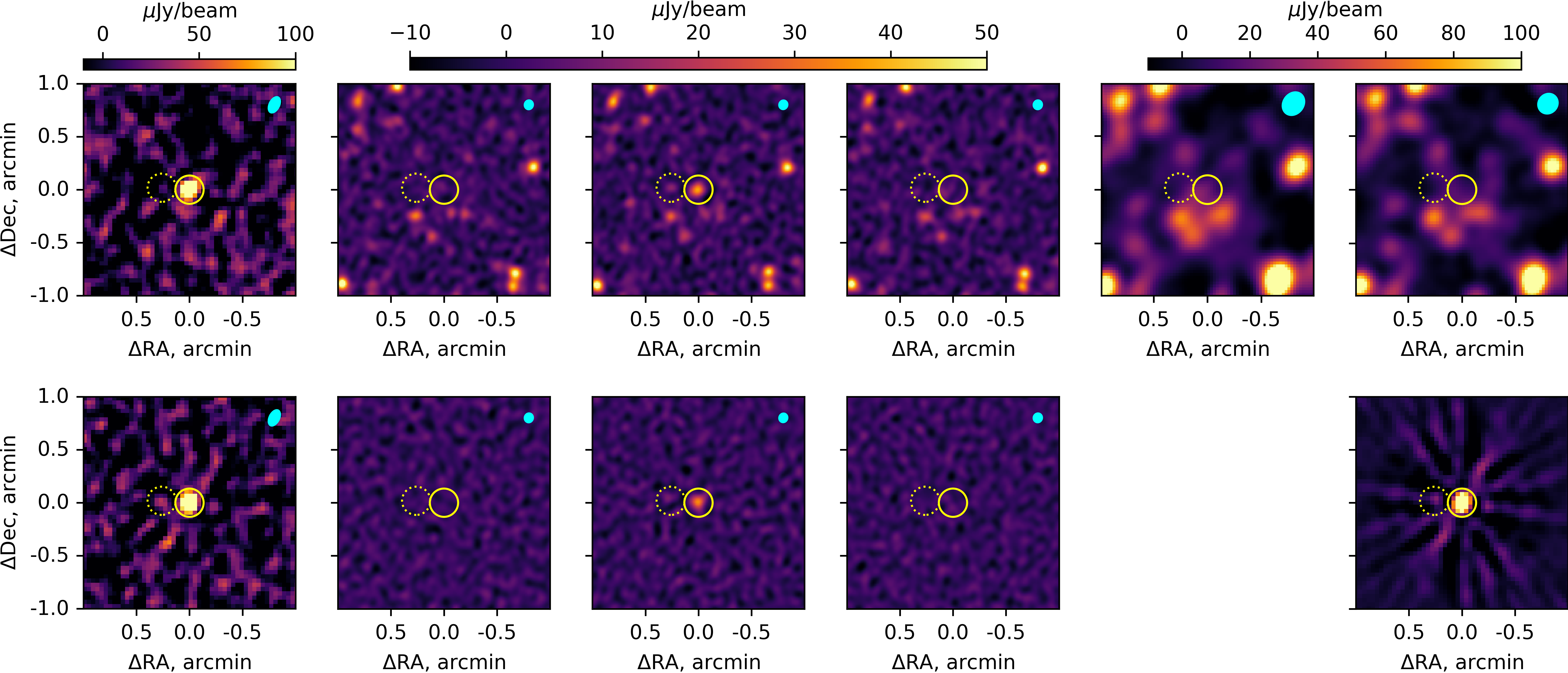}\\
   \caption{\label{fig:l2-cutouts}$2\arcmin\times2\arcmin$ image cutouts centred on Gaia DR3 6865945581361480448. 
   \revone{Top row: Stokes $I$, bottom row: Stokes $V$}. From left to right: (i) L2 epoch, 240~s image at peak of detection (not deconvolved); (ii--iv) deep L-band images for epochs L1, L2 and L4; (v, vi) deep UHF images for epochs U2 and U3 \revone{(Stokes $I$ only). The bottom right image shows the PSF corresponding to the 240~s image at peak of detection.} The position of Gaia DR3 6865945581361480448 is indicated by the solid circle; the dotted circle indicates ``source'' B (see text). 
   }
\end{figure*}

\begin{table}
   \begin{tabular}{cccc}   
     & Stokes $I, \mu\mathrm{Jy}$ & Stokes $V, \mu\mathrm{Jy}$ & $V/I$ \\
    \hline   
    20-Jun-2021 (L2) full &  $70\pm5$	& $64\pm5$	& 92\% \\
    20-Jun-2021 (L2) pre-flare &  $70\pm9$	& $58\pm8$	& 82\%\\
    20-Jun-2021 (L2) flare     &  $491\pm36$	& $425\pm33$	& 87\%\\
    20-Jun-2021 (L2) post-flare &  $27\pm8$?\\
    31-May-2021 (L1) &  $14\pm5$			\\
    27-Jul-2021 (L3) &  $28\pm11$			\\
    20-Jun-2023 (L4) &  $16\pm5$			\\
    27-Jul-2021 (U1) &  $10\pm7$			\\
    8-Aug-2021 (U2) &  $39\pm6$		\\
    4-Feb-2022 (U3) &  $15\pm6$			\\
   \end{tabular}
   \caption{\label{tab:fluxes}Full-band flux measurements for Gaia DR3 6865945581361480448. 
   Fluxes are reported as peak flux over a $6\arcsec$ circular aperture centred on the detected MeerKAT position, \revone{and are corrected for the primary beam response. In the
   case of the flare, flux is measured in the 240~s image at the peak of the light curve.}}
   \end{table}
   
   \begin{figure}
      \includegraphics[width=\columnwidth]{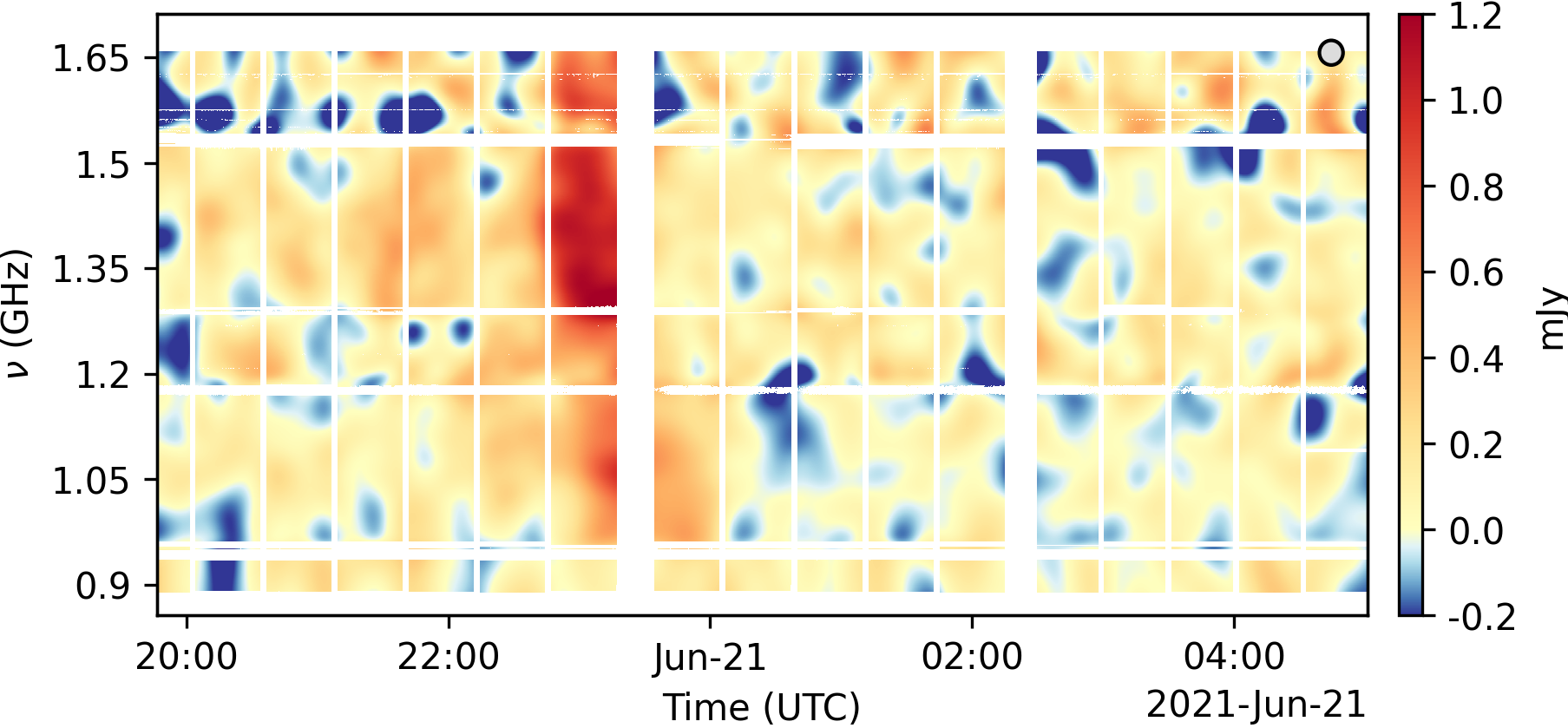}
      \includegraphics[width=\columnwidth]{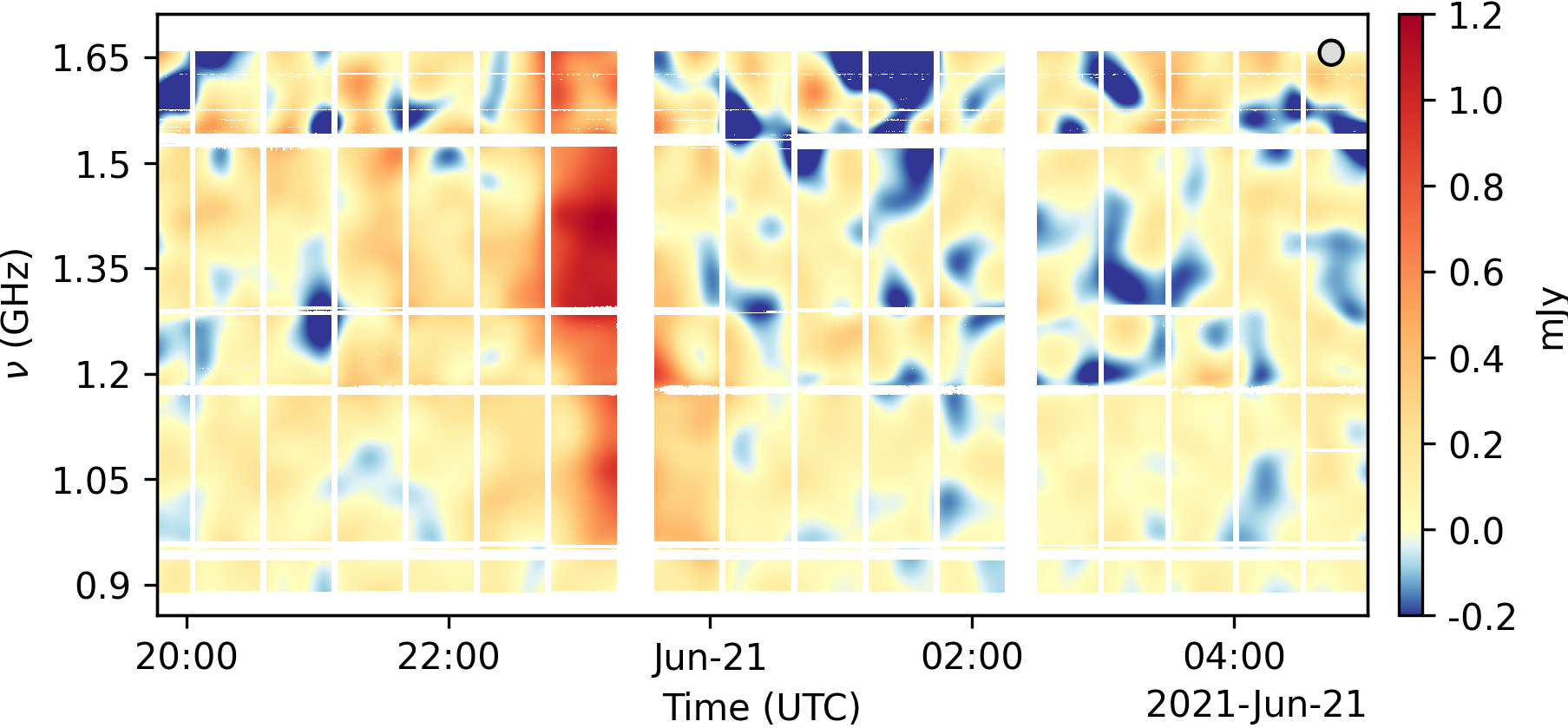}
      \caption{\label{fig:l2-ds}\revone{Primary beam corrected Stokes $I$ (top) and $V$ (bottom) dynamic spectra of Gaia DR3 6865945581361480448, smoothed to 35~MHz and 650~s (FWHM of smoothing kernel is indicated in the top right). Gaps in the dynamic spectra correspond to calibrator scans and fully flagged (RFI-contaminated) bands. Note that the noise across time and frequency is not uniform, due to (a) primary beam correction and (b) the frequency ranges between 0.9--1, 1.2--1.3 and 1.55--1.65 GHz having a larger flagged data fraction.}}
   \end{figure}

The PARROT L2 epoch was processed with sole intent of validating TRON on a known variable source, i.e. the PARROT itself. While the PARROT was reliably detected (at timescales of 15 seconds and longer), the reprocessing yielded a second transient discovery, MKT J200730.4$-$203550, detected at timescales of 30 seconds and longer (Figs.~\ref{fig:l2}, \ref{fig:l2-lc}). \revone{The fitted FK5 position of the radio source is $\alpha=20^\mathrm{h}07^\mathrm{m}30\fs41(1),$ $\delta=-20\degr35\arcmin49\farcs6(2)$. This matches to within 1\farcs5 the position of Gaia DR3 6865945581361480448 \citep{gaia-2016, gaia-2023}. Given the average density of Gaia sources in this part of the sky, the probability 
of a false-positive association within an $r=1\farcs5$ circular region is 1.4\%. We note that arcsecond-level astrometric offsets in MeerKAT radio positions are not unusual \citep[e.g.][]{heywood22-gc}, and are most likely driven by uncertainties in the gain calibrator and the correlator geometric model.} 

\revone{Gaia DR3 6865945581361480448 is also noted in several other stellar catalogues using the VizieR catalogue access tool \citep{vizier2000}, with astrometry, proper proper motion and spectrophotometric estimates that are in agreement with those obtained using Gaia \citep{Monet_2003, zacharias05, federov11, girard11}.} The star has a calculated distance of $1334.38^{+59.36}_{-84.01}\,\text{pc}$ placing it well within the Galaxy as a G type star (T$_{\text{eff}}=5299.121\,\text{K}$, \citealt{gaia-catalogue-2023}).


Having reprocessed the visibility data with the source excluded from the deconvolution mask, we are able to obtain a light curve for the source that is not mean-subtracted (as opposed to the normal TRON outputs, which yield mean-subtracted light curves). This is shown in Fig.~\ref{fig:l2-lc}, in units of apparent flux. The light curve shows a substantial flare between approximately 21:20 and 00:05 UTC, as well as detectable flux outside the flare. 

The source being off-axis, the power beam in its direction will vary in time and frequency. We use the MeerKAT primary beam \revone{holography} measurements of \citet{mdv-beams} to compute the expected power beam towards the source. The power beam gain as a function of time for the bottom (highest gain), mid and top (lowest gain) of the band is also plotted in Fig.~\ref{fig:l2-lc}. The variation of the beam gain in time is not significant for our purposes, being within 2\%, 5\% and 10\% at the bottom, middle and top of the band, respectively. We adopt a mean value of 0.55 for the mid-L-band power beam gain (and 0.8 for UHF), and apply this value to derive the intrinsic fluxes quoted below. We also note that the instrumental leakage from Stokes $I$ to Stokes $V$ due to the primary beam is expected to be below 0.15\% \revone{as per the above measurements} (not plotted).

\revone{The star manifests itself as a $70\pm5\mu\mathrm{Jy}$ continuum source in the deep 10~h image, which corresponds to the mean flux value across the light curve. The flare peaks at $491\pm36\mu\mathrm{Jy}$. To get an estimate for the fluxes pre- and post-flare, we image those segments of the data separately, and fit a Gaussian component. The resulting flux estimates are summarized in Table~\ref{tab:fluxes}. The source is reliably detected pre-flare in Stokes $I$ and $V$. Post-flare, there is only a tentative detection in Stokes $I$ (a $3.5\sigma$ peak, slightly offset from the source position, so we can't exclude the possibility of this being a noise excursion), and no detection in Stokes $V$. The source shows a very high circular polarization fraction of over 85\% both pre- and during the flare. We should note that, given the complex behaviour of the dynamic spectrum (discussed below), these wideband flux measurements are crude at best. We do not detect any linear polarization component, neither in the wideband images, nor in the Stokes $QU$ dynamic spectra, nor in the Faraday spectrum yielded by rotation measure synthesis.}
 
With multiple observations of the field available as part of the PARROT follow-up \citep{parrot}, we are able to check if the source is detected in other epochs. None of them show a clear (over $4\sigma$) detection (Fig.~\ref{fig:l2-cutouts}), however there is a hint of emission in the L4 and U2 epochs, just to the southeast of Gaia DR3 6865945581361480448. Its association with the star is uncertain, particularly in the UHF images, due to blending with nearby confusing sources.  For all other epochs, we may take the peak flux at the source position to be an upper limit. \revone{The L2 image also contains a hint (23$\mu\mathrm{Jy}$ peak) of a compact source, designated ``B'', just to the east of the star, which is also clearly visible in the 240~s Stokes $V$ image. Any physical association is unlikely -- the the on-sky separation between source B and the star would correspond to a minimum physical distance of approximately 21,000 AU -- and the ``source'' is almost certainly associated with a point spread function (PSF) sidelobe (bottom right image of Fig.~\ref{fig:l2-cutouts}) produced at the peak of the flare.
Note that the star is excluded from the deconvolution mask in these images.}

Using the L2 epoch observations, we construct a dynamic spectrum of the source, using 
the {\sc DynSpecMS} tool\footnote{\url{https://github.com/cyriltasse/DynSpecMS}} \revone{\citep{Tasse2025,CRDraconis,parrot,mining1}}. This is presented in Fig.~\ref{fig:l2-ds}, \revone{and shows evidence of drift across the passband of order $\sim$ 10 MHz/min}.

In the Gaia DR3 release, this star is cataloged as having been observed 34 times between September 2014 and May 2017, with a mean G magnitude of 14.7883 ± 0.0021. The associated time-series shows clear evidence of consistent variability, with a $\Delta m$ $\sim$ 0.11, and a Lomb-Scargle analysis yields a dominant period of 1.19 days. Gaia DR3 6865945581361480448 is coincident with star AP60750815 in the All Sky Automated Survey for SuperNovae (ASAS-SN) program \citep{2014ApJ...788...48S, 2017PASP..129j4502K}, with a $m_{V}$ = 15.09 $\pm$ 0.056 over 198 epochs between May 2014 and September 2018, again showing a consistently variable albeit noisier time-series ($\Delta m$ $\sim$ 0.3) from which a Lomb-Scargle analysis recovers the same 1.19 day period. The duration of MKT J200730.4$-$203549's L2 flare ($\sim$ 2 hours), its high circular polarization and associated estimated brightness temperature ($>$ 10$^{12}$K) all point towards a coherent emission process. A supervised machine learning based classification of 12.4 million variable sources in the Gaia DR3 catalog flagged this source as a likely RS CVn variable \citep{2023A&A...674A..14R}.

RS CVn binaries are chromospherically active on account of their tight, tidally locked, orbital orientation resulting in high rotational velocities and complex magnetic field topologies associated with prevalent starspots on one or both of the component stars, which are typically of the F, G or K spectral type. These conditions give rise to distinct emission properties observable at multiple passbands. The orbital period of such systems - ranging from $\sim$ 1 day to $\sim$ 30 days - is easily detectable optically, with a magnitude variation typically of $\pm$ 0.2, and in certain viewing alignments, periodic eclipses are present. The rotationally enhanced magnetic activity manifests itself with the presence of strong H$\alpha$, Ca II, H and K emission lines, with flare-like activity observed in both the EUV/X-ray and radio passbands when presumed magnetic loop reconnection events occur, with the \revone{resulting} observed long duration ($\sim$ hours) and highly circularly polarised radio emission. Recent observations at low frequencies using LOFAR for 14 RS CVn systems suggest the source of this emission is one or both of the stellar chromospheres or via electrodynamic interactions between the stars \citep{2021A&A...654A..21T}.


\revone{Whilst RS CVn systems are known to produce both quiescent and enhanced (`flare') radio emission, the conventional understanding is that gyrosynchrotron processes dominate, yielding low levels of circularly polarised emission \citep{rscvn_consensus}. However, there have also been a number of cases reported in the literature where prolonged ($\sim$ hours) pronounced flare emission from such systems has been reported at GHz bands to be associated with consistently high levels ($>$ 40\%) of circular polarisation associated with brightness temperatures > 10$^{12}$ K with evidence of drift across the bandpasses ($\sim$ 1-10 MHz/min) that are difficult to explain without involving a coherent process, with several authors implicating the electron cyclotron maser instability (ECMI) \citep{mutel87, osten00, garcia-sanchez03, slee08}.}

\revone{Given the bandpass of 856 MHz and a central channel frequency of 1.284 GHz, one can estimate both peak and quiescent luminosity of $L_\mathrm{peak}$ $\sim$ 1.04 $\times$ 10$^{18}$ erg/s/Hz and $L_\mathrm{q}$ $\sim$ 1.49 $\times$ 10$^{17}$ erg/s/Hz. This compares favourably with RS CVn luminosities reported between 10$^{16}$ and 10$^{18}$ erg/s/Hz at this frequency band, with the latter being associated with `superflare' activity having its origin in local coherent emission processes in the RS CVn system \citep{mutel87}}

The observational data we have to hand -- periodicity of 1.19 days, consistent variability of $\sim$ $\pm$ 0.2 magnitudes, and the physical properties of the observed flare -- given its duration $\sim$ 2 hours and high circular polarisation fraction -- are all consistent with Gaia DR3 6865945581361480448/MKT J200730.4$-$203550 being a distant RS CVn. The dynamic spectrum (Fig.~\ref{fig:l2-ds}) is of particular interest, as there is some suggestion of drifting across the 0.9--1.7 GHz bandpass similar to ECMI emission observed from the RS CVn binary HR 1099 at 1.384 GHz albeit with a smaller bandwidth of 104 MHz, again associated with $\sim$ 2 hour highly circularly polarised flare events \citep{2008PASA...25...94S}. An examination of the archived Gaia XP mean spectrum for 6865945581361480448 shows no evidence for emission lines, however this is not unexpected given its low spectral resolution and averaged nature -- follow-up high resolution spectroscopy would likely confirm the characteristic chromospheric emission line signature.

\section{Conclusions}

MeerKAT has proven itself to be a remarkable instrument for the detection of faint, medium-timescale radio transients and variables, and we can expect a systematic reprocessing of archival synthesis imaging data to yield many new discoveries. A reprocessing of the PARROT fields yielded a new transient, associated with the star Gaia DR3 6865945581361480448. \revone{The optical properties of the star are indicative of it being a distant RS CVn binary, 
and the observed radio behaviour (i.e. the nature of the dynamic spectrum, and the high degree of circular polarization) are consistent with an ECMI burst from such a system.}

Unlike Letter I of this series, where the detections were made at the centre of the observed fields, and where the fields were already known to host variable sources, the detection reported on here is truly serendipitous, and underscores the value of the `mining' approach. The detection was made far from the field centre, while the observation was done for unrelated reasons. This shows that the TRON pipeline is capable of discovering transients and variables of different astrophysical nature, over the entire field of view of the telescope.

\paragraph*{Data Availability.} The raw data underlying this article is publicly available via the SARAO archive\footnote{\url{https://archive.sarao.ac.za}}, under proposal IDs SSV-20200715-SA-01.

\paragraph*{Acknowledgements.}  The MeerKAT telescope is operated by the South African Radio Astronomy Observatory, which is a facility of the National Research Foundation, an agency of the Department of Science and Innovation. OMS's, JSK's and VGGS's research is supported by the South African Research Chairs Initiative of the Department of Science and Technology and National Research Foundation (grant No. 81737).  MAT gratefully acknowledges the support of the UK's Science \& Technology Facilities Council (STFC) through grant awards ST/R000905/1 and ST/W00125X/1. We acknowledge the financial support of the Breakthrough Listen project. Breakthrough Listen is managed by the Breakthrough Initiatives, sponsored by the Breakthrough Prize Foundation.

\bibliographystyle{mnras} 
\bibliography{parrot-l2} 
 
\label{lastpage}

\end{document}